\documentclass[aps,prd,nofootinbib,amsmath,amssymb,superscriptaddress,twocolumn,10pt]{revtex4}
\usepackage{graphicx}
\usepackage{dcolumn}
\usepackage{bm}
\usepackage{amssymb}
\usepackage{latexsym}
\usepackage{booktabs}
\usepackage{amsmath}
\usepackage{multirow}
\usepackage[colorlinks=true, linkcolor=red, citecolor=blue]{hyperref}

\newcommand{\be}{\begin{equation}}
\newcommand{\ee}{\end{equation}}
\newcommand{\bq}{\begin{eqnarray}}
\newcommand{\eq}{\end{eqnarray}}

\begin{document}

\title{Testing models of vacuum energy interacting with cold dark matter}

\author{Yun-He Li}
\affiliation{Department of Physics, College of Sciences, Northeastern University, Shenyang
110004, China}
\author{Jing-Fei Zhang}
\affiliation{Department of Physics, College of Sciences, Northeastern University, Shenyang
110004, China}
\author{Xin Zhang\footnote{Corresponding author.}}
\email{zhangxin@mail.neu.edu.cn} \affiliation{Department of Physics, College of Sciences,
Northeastern University, Shenyang 110004, China}
\affiliation{Center for High Energy Physics, Peking University, Beijing 100080, China}

\begin{abstract}

We test the models of vacuum energy interacting with cold dark matter and try to probe the possible deviation from the $\Lambda$CDM model using current observations. We focus on two specific models, $Q=3\beta H\rho_{\Lambda}$ and $Q=3\beta H\rho_c$. The data combinations come from the Planck 2013 data, the baryon acoustic oscillations measurements, the type-Ia supernovae data, the Hubble constant measurement, the redshift space distortions data and the galaxy weak lensing data. For the $Q=3\beta H\rho_c$ model, we find that it can be tightly constrained by all the data combinations, while for the $Q=3\beta H\rho_{\Lambda}$ model, there still exist significant degeneracies between parameters. The tightest constraints for the coupling constant are $\beta=-0.026^{+0.036}_{-0.053}$ (for $Q=3\beta H\rho_{\Lambda}$) and $\beta=-0.00045\pm0.00069$ (for $Q=3\beta H\rho_c$) at the $1\sigma$ level. For all the fit results, we find that the null interaction $\beta=0$ is always consistent with data. Our work completes the discussion on the interacting dark energy model in the recent Planck 2015 papers. Considering this work together with the Planck 2015 results, it is believed that there is no evidence for the models beyond the standard $\Lambda$CDM model from the point of view of possible interaction.

\end{abstract}

 \maketitle


Modern cosmology encounters a great puzzle in explaining the late-time cosmic acceleration, as a universe with barotropic or pressureless fluids, evolving according to the laws of general relativity, cannot fit the observational data \cite{Riess98,Perlmutter98}. This puzzle is phenomenologically solved by introducing a new component with negative pressure, generally called dark energy. However, the fundamental nature of dark energy is still unknown. At present, a cosmological constant $\Lambda$ with equation of state $w=-1$ is the simplest candidate of dark energy, which, however, is plagued with the so-called fine-tuning problem and coincidence problem \cite{14dereview,Carroll01}. These theoretical problems provide cosmologists with the motivation to consider some complex theories for explaining the cosmic acceleration, such as the dynamical dark energy models or the modified gravity (MG) theories. In particular, in order to solve the coincidence problem, cosmologists have widely investigated the interacting dark energy (IDE) scenario \cite{intde1,Comelli:2003cv,Zhang:2005rg,Zhang:2005rj,Cai:2004dk,Guo:2004xx,Amendola:1999qq,TocchiniValentini:2001ty,Chimento:2003iea,Farrar:2003uw,Sadjadi:2006qp,Barrow:2006hia,Zhang:2007uh,Abdalla:2007rd,Guo:2007zk,Bean:2007ny,CalderaCabral:2008bx,Bean:2008ac,Szydlowski:2008by,Chen:2008ft,Li:2009zs,Zhang:2009qa,Gavela:2009cy,Chimento:2009hj,He:2010ta,Cui:2010dr,Li:2010eu,He:2010im,Li:2011ga,Fu:2011ab,Zhang:2012uu,Li:2013bya,Wang:2014oga,Faraoni:2014vra,Xu:2011tsa,Xu:2013jma,Wang:2014iua,Geng:2015ara,Duniya:2015nva,Cai:2009ht,Gavela:2010tm,Martinelli:2010rt,CalderaCabral:2009ja,Koyama:2009gd,Majerotto:2009np,Valiviita:2009nu,Boehmer:2009tk,Clemson:2011an,Zhang:2013lea,Zhang:2013zyn,Ferrer:2004nv,Sasaki:2006kq,Salvatelli:2013wra,Xia:2009zzb,Yang:2014gza,yang:2014vza,Cai:2015zoa,Valiviita:2015dfa}, where dark energy directly interacts with dark matter by exchanging energy and momentum.

Regardless of the theoretical problems of $\Lambda$, the $\Lambda$ cold dark matter ($\Lambda$CDM) model is still the most competitive cosmological model and is now even taken as a prototype of the standard model in cosmology, as it can excellently fit most of the current data with the least free parameters \cite{Ade:2015xua}. Nevertheless, other complex models mentioned above are also not excluded by the observations. Given this fact, we still need to continuously test the validity of the $\Lambda$CDM model. Undoubtedly, any deviation from it, if confirmed by observations, would be a major breakthrough in cosmology. In fact, some degree of tension between different observations has been reported for the $\Lambda$CDM model since the Planck 2013 results were published \cite{Ade:2013zuv}. This tension has attracted much interest from cosmologists. Many works tried to probe or disprove these possible deviations from the standard $\Lambda$CDM model (see, e.g., Refs.~\cite{Zhang:2014dxk,Zhang:2014nta,Li:2014dja,Zhang:2014lfa,Li:2015poa}). Particularly, in recent Planck 2015 results, the Planck Collaboration specifically discussed this topic in Ref.~\cite{Ade:2015rim}, where a series of dynamical dark energy and MG models were constrained by Planck 2015 data in combination with other astrophysical observations. The constraint results showed that a $\Lambda$CDM universe still resides in the allowed region of parameter space at 2$\sigma$ level for most of the data combinations.

Although a number of models were studied in the Planck 2015 results, the IDE models were somehow excluded (only a specific coupled quintessence model was discussed). In fact, compared with the dynamical dark energy and MG models, the IDE models can provide more features to fit the observations. Namely, the IDE models can not only affect the Universe's background evolution but also directly change the growth history. In fact, they are able to both enhance and suppress the structure growth (see Ref.~\cite{CalderaCabral:2009ja}), whereas, generically, the dynamical dark energy models only affect the background evolution (expansion history)\footnote{For a dynamical dark energy model with $w$ a constant or varying slowly with time, it predicts a growth index, $\gamma=0.55+0.05[1+w(z=1)]$, with small deviation from the standard $\Lambda$CDM value 0.55 \cite{Wang:1998gt,Linder:2005in}. Thus, dynamical dark energy models generally do not induce significant new feature to the history of growth of structure.} and the MG models only tend to enhance the structure growth. We notice that a special IDE model with dark energy characterized by a scalar field (namely, a coupled quintessence model) was studied in the Planck 2015 results \cite{Ade:2015rim}. However, unlike a general IDE model where dark energy is phenomenologically described by some fluid, such a specific coupled quintessence model [$m(\phi)=m_0 e^{-\beta\phi}$ and $V(\phi)=V_0\phi^{-\alpha}$] always tends to enhance the structure growth; i.e., the effective gravitational constant, $G_{\rm eff}=G(1+\beta^2)$ \cite{Ade:2015rim}, is always larger than the standard gravitational constant $G$, no matter what sign of coupling constant $\beta$ is chosen. Besides, due to the degeneracy between the coupling constant and the potential slope of the scalar field, the one-dimensional posterior distribution for the coupling constant somehow depends on the prior of the potential slope. (See Figs. 21 and 22 in Ref.~\cite{Ade:2015rim}.) From the above analysis, we believe that the discussion on the IDE model in Ref.~\cite{Ade:2015rim} should be extended.

Our aim is to test if there is any evidence for the deviation from the standard (six-parameter) $\Lambda$CDM model from the point of view of possible interaction between dark sectors. To take careful aim at such a test, we only consider a one-parameter extension to $\Lambda$CDM in this paper. In other words, we focus on a special type of IDE model, i.e., the decaying vacuum energy model (vacuum energy decays to cold dark matter, or vice versa) where dark energy is the vacuum energy with $w=-1$. In these models, the only extra parameter relative to $\Lambda$CDM comes from the coupling between vacuum energy and cold dark matter. Constraining this extra parameter could provide a test for the possible deviation from $\Lambda$CDM. The energy conservation equations for the densities of vacuum energy ($\rho_{\Lambda}$) and cold dark matter ($\rho_c$) satisfy
\begin{align}
&\dot{\rho}_{\Lambda} = Q,\label{eqn:backconsv}\\
&\dot{\rho}_c = -3H\rho_c-Q,\label{eqn:backconsvc}
\end{align}
where a dot denotes the derivative with respect to the cosmic time $t$, $H$ is the Hubble parameter, and $Q$ denotes the energy transfer rate. The decaying vacuum energy models have been widely studied in the literature; see, e.g., Refs. \cite{Zhang:2012sya,Yin:2015pqa,Wang:2014xca,G:2014mea,Chimento:2013rya,Xu:2011qv,Salvatelli:2014zta}.

Since we currently have no fundamental theory to determine the form of the energy transfer rate, we consider two phenomenological forms of $Q$, namely, $Q=3\beta H\rho_{\Lambda}$ and $Q=3\beta H\rho_c$, where $\beta$ denotes the coupling constant. It can be seen that such models only have one more free parameter $\beta$ compared with the standard $\Lambda$CDM model, and $\beta=0$ recovers the standard $\Lambda$CDM universe. Thus, by constraining the value of $\beta$, we can clearly see whether any deviation from the $\Lambda$CDM model exists. Even with one extra free parameter, these IDE models can also change the background and perturbation evolutions of the dark sectors, simultaneously, which can be seen from the above equations and the following discussions.

In the decaying vacuum energy models, if the cosmological perturbations are considered, the vacuum energy should anyhow be disturbed by the perturbation of dark matter. Then the vacuum energy would no longer be a pure background but has perturbation, in principle. In return, the evolution of dark matter perturbation will also be affected by the perturbation of vacuum energy. The above scenario is embodied in the covariant conservation laws, $\nabla_\nu T^{\mu\nu}_I = Q^\mu_I$ for $I=\Lambda$ and $c$, where $T^{\mu\nu}$ denotes the energy-momentum tensor and $Q^\mu$ is the energy-momentum transfer vector. In this work, we choose $Q^\mu_{\Lambda}=-Q^\mu_c=Qu^\mu_c$ with $u^\mu_c$ the four-velocity of cold dark matter, so that there is no momentum transfer in the rest frame of cold dark matter.

Now that the vacuum energy has perturbation, the crucial problem of how to handle this perturbation occurs. Note that we must take this problem seriously, since the calculation of the dark energy perturbation in a conventional way has induced several instabilities. A well-known example is that the perturbation of dark energy will diverge when $w$ crosses the phantom divide $w=-1$ \cite{Vikman:2004dc,Hu:2004kh,Caldwell:2005ai,Zhao:2005vj}. In addition, in an IDE model the cosmological perturbations will blow up on the large scales at specific parameter values \cite{Valiviita:2008iv,Clemson:2011an,He:2008si}.\footnote{The cosmological perturbations are stable only when $w<-1$ for $Q\propto\rho_c$ model, and $w>-1$ and $\beta>0$ for $Q\propto\rho_{de}$ model, if $w$ is close to a constant.} These issues may result from the incorrect calculation of the pressure perturbation of dark energy, as pointed out in Refs.~\cite{Li:2014eha,Li:2014cee}. Under such circumstances, we handle the perturbation of vacuum energy in this work based on the parametrized post-Friedmann (PPF) approach \cite{Hu:2008zd,Fang:2008sn}. The PPF framework applied to the IDE models, completed in Ref.~\cite{Li:2014eha} for the first time, can successfully avoid the large-scale instability of IDE models and help us to probe the full parameter space of $\beta$. For more information about the PPF approach and the calculation of the cosmological perturbations in the IDE models, see Refs.~\cite{Li:2014eha,Li:2014cee}.

To solve the background and perturbation equations for these two decaying vacuum energy models, we modified the public {\tt PPF} \cite{Hu:2008zd,Fang:2008sn} and Boltzmann code {\tt CAMB} \cite{Lewis:1999bs}. We use the public Markov-chain Monte Carlo package {\tt CosmoMC} \cite{Lewis:2002ah} to explore the parameter space. The free parameter vector is $\{\omega_b,\, \omega_c,\, \theta_{\rm{MC}},\, \tau,\,n_{\rm{s}},\, {\rm{ln}}(10^{10}A_{\rm{s}}),\,\beta\}$, where $\omega_b$ and $\omega_c$ are the physical baryon and cold dark matter densities, respectively, $n_{\rm{s}}$ and ${\rm{ln}}(10^{10}A_{\rm{s}})$ denote the spectral index and the amplitude of the primordial scalar perturbation power spectrum, respectively, $\theta_{\rm{MC}}$ is the approximation to the angular size of the sound
horizon at last-scattering time, and $\tau$ is the optical depth to reionization. We set $\beta$ as a prior [$-0.015$, 0.015] for the $Q=3H\beta\rho_c$ model and [$-0.15$, 0.15] for the $Q=3\beta H\rho_{\Lambda}$ model; other free parameter priors are the same as those in Ref.~\cite{Ade:2013zuv}. In our calculations, we always assume two massless and one massive neutrino species with total mass $\sum m_{\nu}=0.06$ eV. In addition, we also set $f_\zeta=0$ and $c_\Gamma=0.4$, in accordance with those in Refs.~\cite{Li:2014eha,Fang:2008sn}, where $f_\zeta$ and $c_\Gamma$ are two parameters used in the PPF approach: $f_\zeta$ is a function relating the momentum density of DE with that of the other components, and $c_\Gamma$ denotes a transition scale below which DE is smooth. (Note that $f_\zeta$ and $c_\Gamma$ can be calibrated in practice~\cite{Li:2014cee}. However, the choices of their values mentioned above are accurate enough for current observations \cite{Fang:2008sn}.)
 
For the observational data, we employ the Planck data in combination with other cosmological probes. Since the Planck 2015 likelihoods are not yet available, we use the data from the Planck 2013 release including the temperature power spectrum data from Planck \cite{Ade:2013zuv} and the polarization power spectrum data from WMAP \cite{wmap9}. We shall use ``Planck'' to denote the above data combination. Besides, we also use the Planck lensing data \cite{Ade:2013tyw} as an additional option. For other observations, we use the following data sets: 
\begin{itemize}
\item BAO: the baryon acoustic oscillations measurements from SDSS at 
$z=0.15$ \cite{Ross:2014qpa}, BOSS at $z=0.32$ and $z=0.57$ \cite{Anderson:2013zyy}, and 6dFGS at $z=0.106$ \cite{Beutler:2011hx}.
\item SNIa: the type-Ia supernovae data from the Joint Light-curve Analysis (JLA) sample \cite{Betoule:2014frx}. 
\item $H_0$: the Hubble constant measurement $H_0=(70.6\pm3.3)\,{\rm km\,s^{-1}\,Mpc^{-1}}$ \cite{Efstathiou:2013via}.
\item RSD: the redshift space distortions data from BOSS at $z=0.57$ \cite{Samushia:2013yga}. 
\item WL: the galaxy weak lensing data from CFHTLenS \cite{Heymans:2013fya}. All these data are utilized in the same way as Ref.~\cite{Ade:2015rim}. Note that when we use the RSD data, we do not use the BAO data at $z=0.57$. For simplicity, we also use ``BSH'' to denote the BAO+SNIa+$H_0$ combination.
\end{itemize}

\begin{figure}[tbp]
\centering 
\includegraphics[width=7cm]{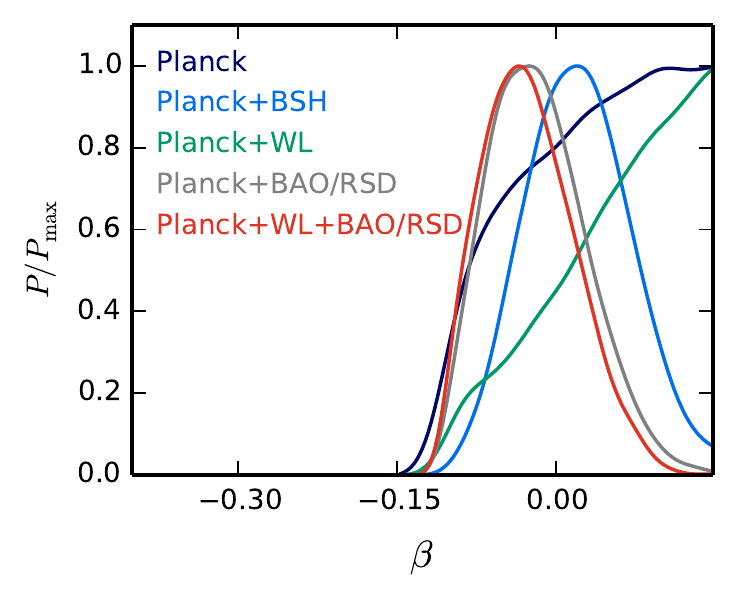}\\
\includegraphics[width=7cm]{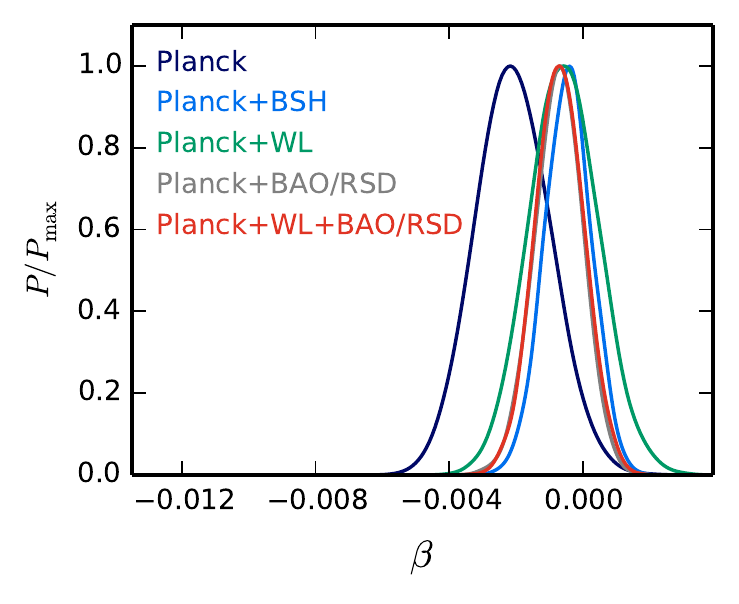}
\hfill
\caption{\label{fig:1d} The one-dimensional posterior distributions for $\beta$. The upper panel corresponds to the $Q=3\beta H\rho_{\Lambda}$ model and the lower panel the $Q=3\beta H\rho_c$ model.}
\end{figure}

\begin{figure*}[tbp]
\centering 
\includegraphics[width=16cm]{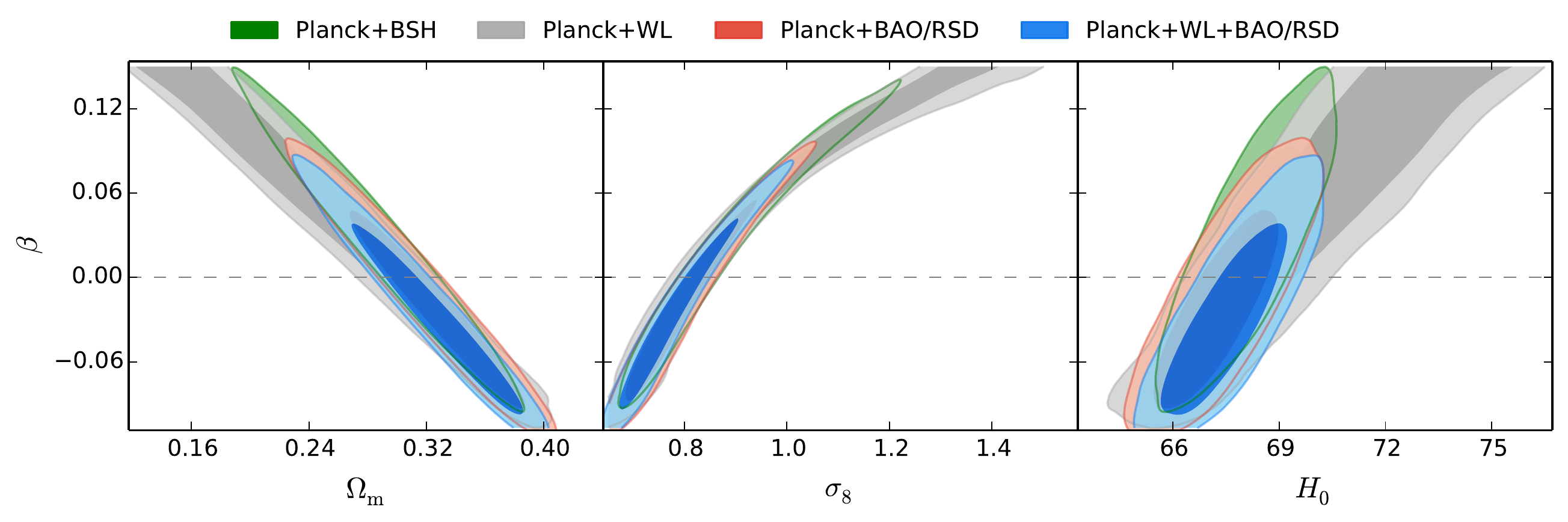}\\
\includegraphics[width=16cm]{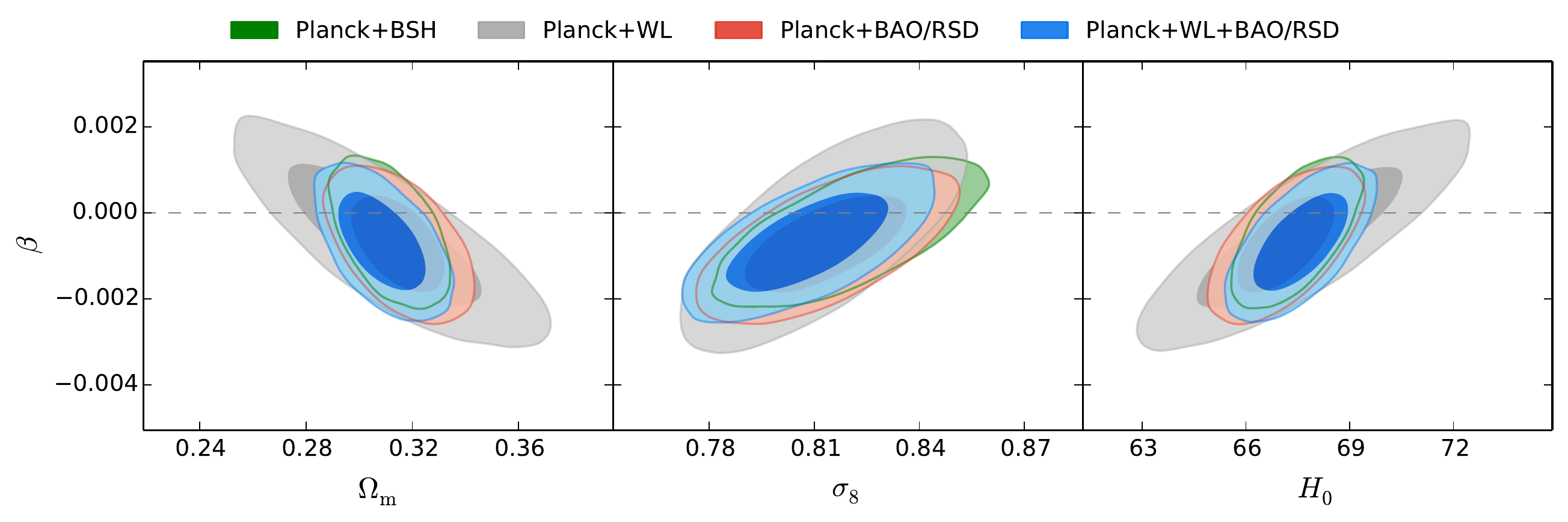}
\hfill
\caption{\label{fig:2d} The two-dimensional marginalized contours (68\% and 95\% CL) for the $Q=3\beta H\rho_{\Lambda}$ model (upper panel) and the $Q=3\beta H\rho_c$ model (lower panel).}
\end{figure*}

\begin{table*}[tbp]
\centering\caption{\label{table1} The mean values and 1$\sigma$ errors for the main parameters of the $Q=3\beta H\rho_{\Lambda}$ model. Note that $\Delta\chi^2_{\rm min}=\chi^2_{{\rm min},\,\Lambda{\rm CDM}}-\chi^2_{\rm min}$, and ``+lensing'' means that the Planck lensing data are combined.}
\begin{tabular}{lccccc}
\hline
\hline
Parameters & Planck  & Planck+BSH &Planck+WL&Planck+BAO/RSD&Planck+WL+BAO/RSD\\
\hline
$\Omega_m$&$0.281^{+0.066}_{-0.094}$&$0.293^{+0.043}_{-0.038}$&$0.242^{+0.047}_{-0.088}$&$0.323^{+0.046}_{-0.032}$&$0.324^{+0.045}_{-0.031}$\\
$\sigma_8$&$0.95^{+0.12}_{-0.25}$&$0.879^{+0.072}_{-0.132}$&$1.01^{+0.16}_{-0.28}$&$0.804^{+0.048}_{-0.103}$&$0.782^{+0.044}_{-0.094}$\\
$H_0$&$68.3^{+2.8}_{-2.6}$&$68.1\pm1.0$&$70.3^{+2.8}_{-2.1}$&$67.4^{+1.0}_{-1.3}$&$67.6^{+1.0}_{-1.2}$\\
$\beta$&$0.036^{+0.114}_{-0.039}$&$0.020^{+0.048}_{-0.053}$&$0.058^{+0.092}_{-0.026}$&$-0.017^{+0.039}_{-0.055}$&$-0.026^{+0.036}_{-0.053}$\\
$\beta$ (+lensing)&$0.031^{+0.110}_{-0.048}$&$0.007^{+0.044}_{-0.050}$&$0.049^{+0.097}_{-0.031}$&$-0.025^{+0.035}_{-0.053}$&$-0.029^{+0.034}_{-0.053}$\\
$\chi^2_{\rm min}$&9804.718&10506.782&9835.890&9808.558&9838.216\\
$\Delta\chi^2_{\rm min}$&0.964&0.118&0.606&0.144&1.94\\
\hline
\hline
\end{tabular}
\end{table*}

\begin{table*}[tbp]
\centering\caption{\label{table2} The mean values and 1$\sigma$ errors for the main parameters of the $Q=3\beta H\rho_c$ model. Note that $\Delta\chi^2_{\rm min}=\chi^2_{{\rm min},\,\Lambda{\rm CDM}}-\chi^2_{\rm min}$, and ``+lensing'' means that the Planck lensing data are combined.}
\begin{tabular}{lccccc}
\hline
\hline
Parameters & Planck  & Planck+BSH &Planck+WL&Planck+BAO/RSD&Planck+WL+BAO/RSD\\
\hline
$\Omega_m$&$0.371^{+0.035}_{-0.039}$&$0.3114^{+0.0086}_{-0.0088}$&$0.31^{+0.023}_{-0.025}$&$0.315\pm0.011$&$0.309^{+0.010}_{-0.011}$\\
$\sigma_8$&$0.809^{+0.015}_{-0.017}$&$0.820\pm0.015$&$0.812\pm0.016$&$0.813\pm0.014$&$0.808\pm0.014$\\
$H_0$&$63.4^{+2.2}_{-2.5}$&$67.4\pm0.7$&$67.6\pm1.9$&$67.1\pm0.9$&$67.6\pm0.9$\\
$\beta$&$-0.0021\pm0.0011$&$-0.00045\pm0.00069$&$-0.0006\pm0.0010$&$-0.00073^{+0.00075}_{-0.00069}$&$-0.00069^{+0.00072}_{-0.00071}$\\
$\beta$ (+lensing)&$-0.0017\pm0.0011$&$-0.00066^{+0.00068}_{-0.00066}$&$-0.00067^{+0.00099}_{-0.00097}$&$-0.00091\pm0.00070$&$-0.00086^{+0.00071}_{-0.00076}$\\
$\chi^2_{\rm min}$&9802.176&10506.964&9836.234&9808.486&9839.048\\
$\Delta\chi^2_{\rm min}$&3.506&$-0.064$&0.262&0.216&1.108\\
\hline
\hline
\end{tabular}
\end{table*}

Now we start to discuss our fit results. First, we constrain the $Q=3\beta H\rho_{\Lambda}$ model. The detailed fitting results are given in Table \ref{table1}. The one-dimensional marginalized posterior distribution curves for $\beta$ are shown in the upper panel of Fig.~\ref{fig:1d}, and the two-dimensional marginalized posterior distribution contours including $\beta$ are shown in the upper panels of Fig.~\ref{fig:2d}. For this model, the Planck data alone and the Planck+WL data combination cannot provide good constraints on the coupling constant. The tightest constraint result comes from the Planck+WL+BAO/RSD data combination, which gives $\beta=-0.026^{+0.036}_{-0.053}$ at $1\sigma$ level. This result slightly favors a negative coupling constant corresponding to vacuum energy decaying into cold dark matter. The Planck+BAO/RSD data provide a similar constraint on $\beta$. While the Planck+BSH data give $\beta=0.020^{+0.048}_{-0.053}$ at the $1\sigma$ level, showing that a positive coupling constant (corresponding to cold dark matter decaying into vacuum energy) is slightly favored. For all the data combinations, we find that there still exist significant degeneracies between parameters in this model.

Next, we constrain the $Q=3\beta H\rho_c$ model. The detailed fitting results are given in Table \ref{table2}. The one-dimensional marginalized posterior distribution curves for $\beta$ are shown in the lower panel of Fig.~\ref{fig:1d}, and the two-dimensional marginalized posterior distribution contours including $\beta$ are shown in the lower panels of Fig.~\ref{fig:2d}. A remarkable feature of this model is that the coupling constant can be tightly constrained for all the data combinations. Even using the Planck data alone, we can also obtain a tight constraint result $\beta=-0.0021\pm0.0011$ at $1\sigma$ level. This phenomenon can be easily understood. Since the interaction starts to work in the early matter-dominated epoch (due to $Q\propto\rho_c$), the time when cosmic microwave background (CMB) is formed, it may induce significant effects on CMB. Another interesting phenomenon is that all the data combinations consistently favor a negative $\beta$, showing that vacuum energy decaying into cold dark matter is more supported. For this model, we find that the tightest constraint comes from the Planck+BSH data combination, which gives $\beta=-0.00045\pm0.00069$ at 1$\sigma$ level.

Finally, we test all the combinations with the additional information from Planck lensing data. The constraint results are slightly improved, as shown in Tables \ref{table1} and \ref{table2}. For a more complete analysis on the contribution of the Planck lensing data to the constraints on the IDE models, see also, e.g., Ref. \cite{Valiviita:2015dfa}. Looking at the constraints with and without the Planck lensing data, we find that the null interaction ($\beta=0$) is still consistent with current observations at the 1$\sigma$ level for most of the data combinations. Our results, together with the Planck 2015 results, show that currently there is no evidence for the models beyond the standard $\Lambda$CDM model. 

We also wish to compare our analysis with previous ones. In our previous works, Refs. \cite{Li:2014eha,Li:2014cee}, we have established the PPF framework for the IDE models and, as examples, we have made brief analyses for the interacting $w$CDM models (for simplicity, we call them the I$w$CDM models, where $w$ is a constant) with $Q=3\beta H\rho_c$ and $Q=3\beta H\rho_\Lambda$. We found that, for the $Q\propto \rho_c$ model, using the geometric measurements can tightly constrain $\beta$ \cite{Li:2014eha}; whereas, for the $Q\propto \rho_\Lambda$ model, using the geometric measurements alone cannot constrain $\beta$ well, and the growth data from RSD can break the degeneracies and improve the constraints significantly \cite{Li:2014cee}. For both I$w$CDM models, we found that the current observations prefer $\beta<0$ at $\sim 1.5\sigma$ level. However, this conclusion is drawn in the models where two extra parameters (i.e., $w$ and $\beta$) relative to $\Lambda$CDM are introduced. Since the current observations can only constrain some combinations of $\beta$ with other parameters---in particular, the degeneracy between $\beta$ and $w$ exists in current data---we actually cannot differentiate the effects from $\beta$ and $w$. Concretely, according to the current observations, $\beta$ is in positive correlation with $w$; i.e., to preserve the observational quantities, decreasing $w$ can be compensated by decreasing $\beta$ \cite{Li:2014eha,Li:2014cee}. Thus, the current data favor simultaneously $w<-1$ and $\beta<0$ \cite{Li:2014eha,Li:2014cee}. In order to avoid such a degeneracy and to test the simplest or minimal extension of $\Lambda$CDM (from the perspective of IDE), we have made an analysis for the interacting $\Lambda$CDM models (referred to as the I$\Lambda$CDM models for simplicity, which are also called the decaying vacuum energy models in this paper) where cold dark matter is allowed to interact with the vacuum energy, without introducing any additional degrees of freedom. Our analysis results show that, when $w$ is fixed at $-1$, $\beta=0$ will be consistent with the current observations at the 1$\sigma$ level. We find that for the I$\Lambda$CDM model with $Q\propto\rho_\Lambda$, $\beta$ still shows strong degeneracies with other parameters, e.g., $\Omega_m$, $\sigma_8$, and $H_0$. Thus, for this model, more accurate RSD measurements are needed in the future to help break the degeneracies (note that in this work, we only use one RSD point, i.e., the BOSS CMASS measurements of the three parameters $D_V/r_{\rm drag}$, $F_{\rm AP}$, and $f\sigma_8$, evaluated at $z_{\rm eff}=0.57$). 

Recently, in Ref. \cite{Salvatelli:2014zta}, the authors investigated the I$\Lambda$CDM model with $Q\propto H\rho_\Lambda$, where the coupling strength $\beta$ is set to be redshift dependent and is taken to be different values in four redshift bins in the analysis. They used the Planck 2013 data in combination with the SN Union2.1 sample and the RSD measurements (10 points) to constrain the model and found that the null interaction ($\beta=0$) is excluded at the $99\%$ confidence level. One of the motivations of this paper is to check whether the null interaction is excluded for the $Q\propto H\rho_\Lambda$ I$\Lambda$CDM model. But our results show that the null interaction case of $\beta=0$ is still consistent with the current observations at the 1$\sigma$ level (except for Planck+WL). The difference between our results and those of Ref. \cite{Salvatelli:2014zta} may come from the following aspects: (i) We only used one RSD point (BOSS CMASS measurement at $z_{\rm eff}=0.57$), but Ref. \cite{Salvatelli:2014zta} used 10 RSD points. We also included the Alcock-Paczynski (AP) effect \cite{Alcock:79} in the RSD measurement. Our utilization of the astrophysical observations is in exact accordance with the Planck Collaboration \cite{Ade:2015rim}. (ii) We handled the perturbation of the vacuum energy within the PPF framework \cite{Li:2014eha,Li:2014cee}, but in Ref. \cite{Salvatelli:2014zta} the perturbation of vacuum energy was neglected. (iii) We assumed that the coupling strength $\beta$ is a constant, but Ref. \cite{Salvatelli:2014zta} assumed $\beta(z)$ is a binned (stepwise-defined) function. It is of great interest to carefully test these aspects, and we leave such an analysis to a future work.

Testing the validity of the standard $\Lambda$CDM model from observations is a vital task in current cosmology. Among the models beyond the $\Lambda$CDM model, the interacting dark energy model can provide more features to fit the observations. In this work, we consider the simplest or minimal extension of $\Lambda$CDM from the perspective of interacting dark energy, i.e., the decaying vacuum energy models (or the I$\Lambda$CDM models).
We have studied two decaying vacuum energy models with the energy transfer $Q=3\beta H\rho_{\Lambda}$ and $Q=3\beta H\rho_c$, respectively. Both models only contain one more free parameter $\beta$ compared with the $\Lambda$CDM model. By constraining $\beta$, we can clearly see whether any deviation from the $\Lambda$CDM model exists. We used the PPF approach to calculate the perturbation of vacuum energy. Our data combinations come from the Planck data, the BAO measurements, the SNIa data, the $H_0$ measurement, the RSD data and the WL data. For the $Q=3\beta H\rho_c$ model, we found that it can be tightly constrained for all the data combinations, while for the $Q=3\beta H\rho_{\Lambda}$ model, there still exist significant degeneracies between parameters. The tightest constraints for $\beta$ are $\beta=-0.026^{+0.036}_{-0.053}$ (for $Q=3\beta H\rho_{\Lambda}$) and $\beta=-0.00045\pm0.00069$ (for $Q=3\beta H\rho_{c}$) at the $1\sigma$ level. For all the fit results, we did not find any evidence beyond the standard $\Lambda$CDM model.


\begin{acknowledgments}
This work was supported by the National Natural Science Foundation of
China under Grants No.~11175042 and No.~11522540, the Provincial Department of Education of
Liaoning under Grant No.~L2012087, and the Fundamental Research Funds for the
Central Universities under Grants No.~N140505002, No.~N140506002, No.~N140504007, and No.~L1505007.
\end{acknowledgments}


\begin{thebibliography}{99}

\bibitem{Riess98}
  A.~G.~Riess {\it et al.}  [Supernova Search Team Collaboration],
  Astron.\ J.\  {\bf 116}, 1009 (1998)
  [astro-ph/9805201].


\bibitem{Perlmutter98}
  S.~Perlmutter {\it et al.}  [Supernova Cosmology Project Collaboration],
  Astrophys.\ J.\  {\bf 517}, 565 (1999)
  [astro-ph/9812133].


\bibitem{14dereview}
S.~Weinberg,
  Rev.\ Mod.\ Phys.\  {\bf 61} (1989) 1.

\bibitem{Carroll01}
S.~M.~Carroll,
  Living Rev.\ Rel.\  {\bf 4} (2001) 1
  [arXiv:astro-ph/0004075].


 \bibitem{intde1}
  L.~Amendola,
  Phys.\ Rev.\  D {\bf 62}, 043511 (2000)
  [arXiv:astro-ph/9908023].
\bibitem{Comelli:2003cv}
  D.~Comelli, M.~Pietroni and A.~Riotto,
  Phys.\ Lett.\  B {\bf 571}, 115 (2003)
  [arXiv:hep-ph/0302080]
\bibitem{Zhang:2005rg}
  X.~Zhang,
  Mod.\ Phys.\ Lett.\  A {\bf 20}, 2575 (2005)
  [arXiv:astro-ph/0503072].

\bibitem{Zhang:2005rj} 
  X.~Zhang,
  Phys.\ Lett.\ B {\bf 611}, 1 (2005)
  [astro-ph/0503075].



\bibitem{Cai:2004dk}
  R.~G.~Cai and A.~Wang,
  JCAP {\bf 0503}, 002 (2005)
  [arXiv:hep-th/0411025].

\bibitem{Guo:2004xx}
  Z.~K.~Guo, R.~G.~Cai and Y.~Z.~Zhang,
  JCAP {\bf 0505}, 002 (2005)
  [astro-ph/0412624].

\bibitem{Amendola:1999qq} 
  L.~Amendola,
  Phys.\ Rev.\ D {\bf 60}, 043501 (1999)
  [astro-ph/9904120].
\bibitem{TocchiniValentini:2001ty} 
  D.~Tocchini-Valentini and L.~Amendola,
  Phys.\ Rev.\ D {\bf 65}, 063508 (2002)
  [astro-ph/0108143].
  
\bibitem{Chimento:2003iea} 
  L.~P.~Chimento, A.~S.~Jakubi, D.~Pavon and W.~Zimdahl,
  Phys.\ Rev.\ D {\bf 67}, 083513 (2003)
  [astro-ph/0303145].
  
\bibitem{Farrar:2003uw} 
  G.~R.~Farrar and P.~J.~E.~Peebles,
  Astrophys.\ J.\  {\bf 604}, 1 (2004)
  [astro-ph/0307316].
\bibitem{Sadjadi:2006qp} 
  H.~M.~Sadjadi and M.~Alimohammadi,
  Phys.\ Rev.\ D {\bf 74}, 103007 (2006)
  [gr-qc/0610080].
\bibitem{Barrow:2006hia} 
  J.~D.~Barrow and T.~Clifton,
  Phys.\ Rev.\ D {\bf 73}, 103520 (2006)
  [gr-qc/0604063].
  
\bibitem{Zhang:2007uh} 
  J.~Zhang, H.~Liu and X.~Zhang,
  Phys.\ Lett.\ B {\bf 659}, 26 (2008)
  [arXiv:0705.4145 [astro-ph]].
  
\bibitem{Abdalla:2007rd} 
  E.~Abdalla, L.~R.~W.~Abramo, L.~Sodre, Jr. and B.~Wang,
  Phys.\ Lett.\ B {\bf 673}, 107 (2009)
  [arXiv:0710.1198 [astro-ph]].
\bibitem{Guo:2007zk} 
  Z.~K.~Guo, N.~Ohta and S.~Tsujikawa,
  Phys.\ Rev.\ D {\bf 76}, 023508 (2007)
  [astro-ph/0702015 [ASTRO-PH]].
\bibitem{Bean:2007ny} 
  R.~Bean, E.~E.~Flanagan and M.~Trodden,
  Phys.\ Rev.\ D {\bf 78}, 023009 (2008)
  [arXiv:0709.1128 [astro-ph]].
\bibitem{CalderaCabral:2008bx} 
  G.~Caldera-Cabral, R.~Maartens and L.~A.~Urena-Lopez,
  Phys.\ Rev.\ D {\bf 79}, 063518 (2009)
  [arXiv:0812.1827 [gr-qc]].
\bibitem{Bean:2008ac} 
  R.~Bean, E.~E.~Flanagan, I.~Laszlo and M.~Trodden,
  Phys.\ Rev.\ D {\bf 78}, 123514 (2008)
  [arXiv:0808.1105 [astro-ph]].
\bibitem{Szydlowski:2008by} 
  M.~Szydlowski, A.~Krawiec, A.~Kurek and M.~Kamionka,
  Eur.\ Phys.\ J.\ C {\bf 75}, no. 99, 5 (2015)
  [arXiv:0801.0638 [astro-ph]].
\bibitem{Chen:2008ft} 
  X.~m.~Chen, Y.~g.~Gong and E.~N.~Saridakis,
  JCAP {\bf 0904}, 001 (2009)
  [arXiv:0812.1117 [gr-qc]].

\bibitem{Li:2009zs} 
  M.~Li, X.~D.~Li, S.~Wang, Y.~Wang and X.~Zhang,
  JCAP {\bf 0912}, 014 (2009)
  [arXiv:0910.3855 [astro-ph.CO]].

\bibitem{Zhang:2009qa} 
  L.~Zhang, J.~Cui, J.~Zhang and X.~Zhang,
  Int.\ J.\ Mod.\ Phys.\ D {\bf 19}, 21 (2010)
  [arXiv:0911.2838 [astro-ph.CO]].

\bibitem{Gavela:2009cy} 
  M.~B.~Gavela, D.~Hernandez, L.~Lopez Honorez, O.~Mena and S.~Rigolin,
  JCAP {\bf 0907}, 034 (2009)
  [JCAP {\bf 1005}, E01 (2010)]
  [arXiv:0901.1611 [astro-ph.CO]].
\bibitem{Chimento:2009hj} 
  L.~P.~Chimento,
  Phys.\ Rev.\ D {\bf 81}, 043525 (2010)
  [arXiv:0911.5687 [astro-ph.CO]].
\bibitem{He:2010ta} 
  J.~H.~He, B.~Wang, E.~Abdalla and D.~Pavon,
  JCAP {\bf 1012}, 022 (2010)
  [arXiv:1001.0079 [gr-qc]].

\bibitem{Cui:2010dr} 
  J.~Cui and X.~Zhang,
  Phys.\ Lett.\ B {\bf 690}, 233 (2010)
  [arXiv:1005.3587 [astro-ph.CO]].


\bibitem{Li:2010eu} 
  B.~Li and J.~D.~Barrow,
  Mon.\ Not.\ Roy.\ Astron.\ Soc.\  {\bf 413}, 262 (2011)
  [arXiv:1010.3748 [astro-ph.CO]].
\bibitem{He:2010im} 
  J.~H.~He, B.~Wang and E.~Abdalla,
  Phys.\ Rev.\ D {\bf 83}, 063515 (2011)
  [arXiv:1012.3904 [astro-ph.CO]].

\bibitem{Li:2011ga} 
  Y.~H.~Li and X.~Zhang,
  Eur.\ Phys.\ J.\ C {\bf 71}, 1700 (2011)
  [arXiv:1103.3185 [astro-ph.CO]].
\bibitem{Fu:2011ab} 
  T.~F.~Fu, J.~F.~Zhang, J.~Q.~Chen and X.~Zhang,
  Eur.\ Phys.\ J.\ C {\bf 72}, 1932 (2012)
  [arXiv:1112.2350 [astro-ph.CO]].

\bibitem{Zhang:2012uu} 
  Z.~Zhang, S.~Li, X.~D.~Li, X.~Zhang and M.~Li,
  JCAP {\bf 1206}, 009 (2012)
  [arXiv:1204.6135 [astro-ph.CO]].

\bibitem{Li:2013bya} 
  Y.~H.~Li and X.~Zhang,
  Phys.\ Rev.\ D {\bf 89}, no. 8, 083009 (2014)
  [arXiv:1312.6328 [astro-ph.CO]].

\bibitem{Wang:2014oga} 
  S.~Wang, Y.~Z.~Wang, J.~J.~Geng and X.~Zhang,
  Eur.\ Phys.\ J.\ C {\bf 74}, no. 11, 3148 (2014)
  [arXiv:1406.0072 [astro-ph.CO]].

\bibitem{Faraoni:2014vra} 
  V.~Faraoni, J.~B.~Dent and E.~N.~Saridakis,
  Phys.\ Rev.\ D {\bf 90}, no. 6, 063510 (2014)
  [arXiv:1405.7288 [gr-qc]].
\bibitem{Xu:2011tsa} 
  X.~D.~Xu and B.~Wang,
  Phys.\ Lett.\ B {\bf 701}, 513 (2011)
  [arXiv:1103.2632 [astro-ph.CO]].
\bibitem{Xu:2013jma} 
  X.~D.~Xu, B.~Wang, P.~Zhang and F.~Atrio-Barandela,
  JCAP {\bf 1312}, 001 (2013)
  [arXiv:1308.1475 [astro-ph.CO]].
\bibitem{Wang:2014iua} 
  J.~S.~Wang and F.~Y.~Wang,
  Astron.\ Astrophys.\  {\bf 564}, A137 (2014)
  [arXiv:1403.4318 [astro-ph.CO]].
\bibitem{Geng:2015ara} 
  J.~J.~Geng, Y.~H.~Li, J.~F.~Zhang and X.~Zhang,
  Eur.\ Phys.\ J.\ C {\bf 75}, no. 8, 356 (2015)
  [arXiv:1501.03874 [astro-ph.CO]].
\bibitem{Duniya:2015nva} 
  D.~G.~A.~Duniya, D.~Bertacca and R.~Maartens,
  Phys.\ Rev.\ D {\bf 91}, 063530 (2015)
  [arXiv:1502.06424 [astro-ph.CO]].
\bibitem{Cai:2009ht} 
  R.~G.~Cai and Q.~Su,
  Phys.\ Rev.\ D {\bf 81}, 103514 (2010)
  [arXiv:0912.1943 [astro-ph.CO]].
\bibitem{Gavela:2010tm} 
  M.~B.~Gavela, L.~Lopez Honorez, O.~Mena and S.~Rigolin,
  JCAP {\bf 1011}, 044 (2010)
  [arXiv:1005.0295 [astro-ph.CO]].
\bibitem{Martinelli:2010rt} 
  M.~Martinelli, L.~Lopez Honorez, A.~Melchiorri and O.~Mena,
  Phys.\ Rev.\ D {\bf 81}, 103534 (2010)
  [arXiv:1004.2410 [astro-ph.CO]].
  
\bibitem{CalderaCabral:2009ja} 
  G.~Caldera-Cabral, R.~Maartens and B.~M.~Schaefer,
  JCAP {\bf 0907}, 027 (2009)
  [arXiv:0905.0492 [astro-ph.CO]].
\bibitem{Koyama:2009gd} 
  K.~Koyama, R.~Maartens and Y.~S.~Song,
  JCAP {\bf 0910}, 017 (2009)
  [arXiv:0907.2126 [astro-ph.CO]].
\bibitem{Majerotto:2009np} 
  E.~Majerotto, J.~Valiviita and R.~Maartens,
  Mon.\ Not.\ Roy.\ Astron.\ Soc.\  {\bf 402}, 2344 (2010)
  [arXiv:0907.4981 [astro-ph.CO]].
\bibitem{Valiviita:2009nu} 
  J.~Valiviita, R.~Maartens and E.~Majerotto,
  Mon.\ Not.\ Roy.\ Astron.\ Soc.\  {\bf 402}, 2355 (2010)
  [arXiv:0907.4987 [astro-ph.CO]].
\bibitem{Boehmer:2009tk} 
  C.~G.~Boehmer, G.~Caldera-Cabral, N.~Chan, R.~Lazkoz and R.~Maartens,
  Phys.\ Rev.\ D {\bf 81}, 083003 (2010)
  [arXiv:0911.3089 [gr-qc]].
\bibitem{Clemson:2011an} 
  T.~Clemson, K.~Koyama, G.~B.~Zhao, R.~Maartens and J.~Valiviita,
  Phys.\ Rev.\ D {\bf 85}, 043007 (2012)
  [arXiv:1109.6234 [astro-ph.CO]].
\bibitem{Zhang:2013lea} 
  J.~Zhang, L.~Zhao and X.~Zhang,
  Sci.\ China Phys.\ Mech.\ Astron.\  {\bf 57}, 387 (2014)
  [arXiv:1306.1289 [astro-ph.CO]].
\bibitem{Zhang:2013zyn} 
  M.~J.~Zhang and W.~B.~Liu,
  Eur.\ Phys.\ J.\ C {\bf 74}, 2863 (2014)
  [arXiv:1312.0224 [astro-ph.CO]].
\bibitem{Ferrer:2004nv} 
  F.~Ferrer, S.~Rasanen and J.~Valiviita,
  JCAP {\bf 0410}, 010 (2004)
  [astro-ph/0407300].
\bibitem{Sasaki:2006kq} 
  M.~Sasaki, J.~Valiviita and D.~Wands,
  Phys.\ Rev.\ D {\bf 74}, 103003 (2006)
  [astro-ph/0607627].
\bibitem{Salvatelli:2013wra} 
  V.~Salvatelli, A.~Marchini, L.~Lopez-Honorez and O.~Mena,
  Phys.\ Rev.\ D {\bf 88}, no. 2, 023531 (2013)
  [arXiv:1304.7119 [astro-ph.CO]].

\bibitem{Xia:2009zzb} 
  J.~Q.~Xia,
  Phys.\ Rev.\ D {\bf 80}, 103514 (2009)
  [arXiv:0911.4820 [astro-ph.CO]].

\bibitem{Yang:2014gza} 
  W.~Yang and L.~Xu,
  Phys.\ Rev.\ D {\bf 89}, no. 8, 083517 (2014)
  [arXiv:1401.1286 [astro-ph.CO]].
\bibitem{yang:2014vza} 
  W.~Yang and L.~Xu,
  JCAP {\bf 1408}, 034 (2014)
  [arXiv:1401.5177 [astro-ph.CO]].
 
\bibitem{Cai:2015zoa} 
  T.~Yang, Z.~K.~Guo and R.~G.~Cai,
  Phys.\ Rev.\ D {\bf 91}, no. 12, 123533 (2015)
  doi:10.1103/PhysRevD.91.123533
  [arXiv:1505.04443 [astro-ph.CO]].
 
      
\bibitem{Valiviita:2015dfa} 
  J.~Valiviita and E.~Palmgren,
  JCAP {\bf 1507}, 015 (2015)
  [arXiv:1504.02464 [astro-ph.CO]].
  



\bibitem{Ade:2015xua}
  P.~A.~R.~Ade {\it et al.}  [Planck Collaboration],
  arXiv:1502.01589 [astro-ph.CO].

\bibitem{Ade:2013zuv}
  P.~A.~R.~Ade {\it et al.}  [Planck Collaboration],
  Astron.\ Astrophys.\  {\bf 571}, A16 (2014)
  [arXiv:1303.5076 [astro-ph.CO]].


\bibitem{Zhang:2014dxk}
  J.~F.~Zhang, Y.~H.~Li and X.~Zhang,
  Phys.\ Lett.\ B {\bf 740}, 359 (2015)
  [arXiv:1403.7028 [astro-ph.CO]].
\bibitem{Zhang:2014nta}
  J.~F.~Zhang, Y.~H.~Li and X.~Zhang,
  Eur.\ Phys.\ J.\ C {\bf 74}, 2954 (2014)
  [arXiv:1404.3598 [astro-ph.CO]].
\bibitem{Li:2014dja}
  Y.~H.~Li, J.~F.~Zhang and X.~Zhang,
  Sci.\ China Phys.\ Mech.\ Astron.\  {\bf 57}, 1455 (2014)
  [arXiv:1405.0570 [astro-ph.CO]].
\bibitem{Zhang:2014lfa}
  J.~F.~Zhang, Y.~H.~Li and X.~Zhang,
  Phys.\ Lett.\ B {\bf 739}, 102 (2014)
  [arXiv:1408.4603 [astro-ph.CO]].
\bibitem{Li:2015poa}
  Y.~H.~Li, J.~F.~Zhang and X.~Zhang,
  Phys.\ Lett.\ B {\bf 744}, 213 (2015)
  [arXiv:1502.01136 [astro-ph.CO]].


\bibitem{Ade:2015rim}
  P.~A.~R.~Ade {\it et al.}  [Planck Collaboration],
  arXiv:1502.01590 [astro-ph.CO].

\bibitem{Wang:1998gt} 
  L.~M.~Wang and P.~J.~Steinhardt,
  Astrophys.\ J.\  {\bf 508}, 483 (1998)
  [astro-ph/9804015].
  
\bibitem{Linder:2005in} 
  E.~V.~Linder,
  Phys.\ Rev.\ D {\bf 72}, 043529 (2005)
  [astro-ph/0507263].
  
\bibitem{Zhang:2012sya} 
  J.~F.~Zhang, Y.~Y.~Li, Y.~Liu, S.~Zou and X.~Zhang,
  Eur.\ Phys.\ J.\ C {\bf 72}, 2077 (2012)
  [arXiv:1205.2972 [astro-ph.CO]].
\bibitem{Yin:2015pqa} 
  J.~L.~Cui, L.~Yin, L.~F.~Wang, Y.~H.~Li and X.~Zhang,
  JCAP {\bf 1509}, no. 09, 024 (2015)
  [arXiv:1503.08948 [astro-ph.CO]].

\bibitem{Wang:2014xca} 
  Y.~Wang, D.~Wands, G.~B.~Zhao and L.~Xu,
  Phys.\ Rev.\ D {\bf 90}, no. 2, 023502 (2014)
  [arXiv:1404.5706 [astro-ph.CO]].
\bibitem{G:2014mea} 
  I.~E.~Sanchez G.,
  Gen.\ Rel.\ Grav.\  {\bf 46}, 1769 (2014)
  [arXiv:1405.1291 [gr-qc]].
\bibitem{Chimento:2013rya} 
  L.~P.~Chimento, M.~G.~Richarte and I.~E.~Sánchez García,
  Phys.\ Rev.\ D {\bf 88}, 087301 (2013)
  [arXiv:1310.5335 [gr-qc]].
\bibitem{Xu:2011qv} 
  L.~Xu, Y.~Wang, M.~Tong and H.~Noh,
  Phys.\ Rev.\ D {\bf 84}, 123004 (2011)
  [arXiv:1112.5216 [astro-ph.CO]].
\bibitem{Salvatelli:2014zta} 
  V.~Salvatelli, N.~Said, M.~Bruni, A.~Melchiorri and D.~Wands,
  Phys.\ Rev.\ Lett.\  {\bf 113}, no. 18, 181301 (2014)
  [arXiv:1406.7297 [astro-ph.CO]].



\bibitem{Vikman:2004dc}
  A.~Vikman,
  Phys.\ Rev.\ D {\bf 71}, 023515 (2005)  [astro-ph/0407107].  
\bibitem{Hu:2004kh}
  W.~Hu,
  Phys.\ Rev.\ D {\bf 71}, 047301 (2005)  [astro-ph/0410680].  
\bibitem{Caldwell:2005ai}
  R.~R.~Caldwell and M.~Doran,
  Phys.\ Rev.\ D {\bf 72}, 043527 (2005)  [astro-ph/0501104].  


\bibitem{Zhao:2005vj}
  G.~B.~Zhao, J.~Q.~Xia, M.~Li, B.~Feng and X.~M.~Zhang,
  Phys.\ Rev.\ D {\bf 72}, 123515 (2005)
  [astro-ph/0507482].

\bibitem{Valiviita:2008iv}
  J.~Valiviita, E.~Majerotto and R.~Maartens,
  JCAP {\bf 0807}, 020 (2008)
  [arXiv:0804.0232 [astro-ph]].


\bibitem{He:2008si}
  J.~-H.~He, B.~Wang and E.~Abdalla,
  Phys.\ Lett.\ B {\bf 671}, 139 (2009)
  [arXiv:0807.3471 [gr-qc]].

\bibitem{Li:2014eha}
  Y.~H.~Li, J.~F.~Zhang and X.~Zhang,
  Phys.\ Rev.\ D {\bf 90}, no. 6, 063005 (2014)
  [arXiv:1404.5220 [astro-ph.CO]].


\bibitem{Li:2014cee}
  Y.~H.~Li, J.~F.~Zhang and X.~Zhang,
  Phys.\ Rev.\ D {\bf 90}, no. 12, 123007 (2014)
  [arXiv:1409.7205 [astro-ph.CO]].

\bibitem{Hu:2008zd}
  W.~Hu,
  Phys.\ Rev.\ D {\bf 77}, 103524 (2008)  [arXiv:0801.2433 [astro-ph]].  
\bibitem{Fang:2008sn}
  W.~Fang, W.~Hu and A.~Lewis,
  Phys.\ Rev.\ D {\bf 78}, 087303 (2008)  [arXiv:0808.3125 [astro-ph]].  

  \bibitem{Lewis:1999bs}
  A.~Lewis, A.~Challinor and A.~Lasenby,
  Astrophys.\ J.\  {\bf 538}, 473 (2000).
  [astro-ph/9911177].

\bibitem{Lewis:2002ah}
  A.~Lewis and S.~Bridle,
  Phys.\ Rev.\ D {\bf 66}, 103511 (2002)  [astro-ph/0205436].  

\bibitem{wmap9}
  G.~Hinshaw {\it et al.}  [WMAP Collaboration],
  Astrophys.\ J.\ Suppl.\  {\bf 208}, 19 (2013)
  [arXiv:1212.5226 [astro-ph.CO]].

\bibitem{Ade:2013tyw}
  P.~A.~R.~Ade {\it et al.}  [Planck Collaboration],
  Astron.\ Astrophys.\  {\bf 571}, A17 (2014)
  [arXiv:1303.5077 [astro-ph.CO]].

\bibitem{Ross:2014qpa}
  A.~J.~Ross, L.~Samushia, C.~Howlett, W.~J.~Percival, A.~Burden and M.~Manera,
  Mon.\ Not.\ Roy.\ Astron.\ Soc.\  {\bf 449}, 835 (2015)
  [arXiv:1409.3242 [astro-ph.CO]].

\bibitem{Anderson:2013zyy}
  L.~Anderson {\it et al.}  [BOSS Collaboration],
  Mon.\ Not.\ Roy.\ Astron.\ Soc.\  {\bf 441}, no. 1, 24 (2014)
  [arXiv:1312.4877 [astro-ph.CO]].

\bibitem{Beutler:2011hx}
  F.~Beutler, C.~Blake, M.~Colless, D.~H.~Jones, L.~Staveley-Smith, L.~Campbell, Q.~Parker and W.~Saunders {\it et al.},
  Mon.\ Not.\ Roy.\ Astron.\ Soc.\  {\bf 416}, 3017 (2011)
  [arXiv:1106.3366 [astro-ph.CO]].

\bibitem{Betoule:2014frx}
  M.~Betoule {\it et al.}  [SDSS Collaboration],
  Astron.\ Astrophys.\  {\bf 568}, A22 (2014)
  [arXiv:1401.4064 [astro-ph.CO]].

\bibitem{Efstathiou:2013via}
  G.~Efstathiou,
  Mon.\ Not.\ Roy.\ Astron.\ Soc.\  {\bf 440}, no. 2, 1138 (2014)
  [arXiv:1311.3461 [astro-ph.CO]].

\bibitem{Samushia:2013yga}
  L.~Samushia, B.~A.~Reid, M.~White, W.~J.~Percival, A.~J.~Cuesta, G.~B.~Zhao, A.~J.~Ross and M.~Manera {\it et al.},
  Mon.\ Not.\ Roy.\ Astron.\ Soc.\  {\bf 439}, no. 4, 3504 (2014)
  [arXiv:1312.4899 [astro-ph.CO]].

\bibitem{Heymans:2013fya}
  C.~Heymans, E.~Grocutt, A.~Heavens, M.~Kilbinger, T.~D.~Kitching, F.~Simpson, J.~Benjamin and T.~Erben {\it et al.},
  Mon.\ Not.\ Roy.\ Astron.\ Soc.\  {\bf 432}, 2433 (2013)
  [arXiv:1303.1808 [astro-ph.CO]].

\bibitem{Alcock:79}
C.~Alcock and B.~Paczynski, 
 \nat, {\bf 281}, 358 (1979)
\end{thebibliography}
\end{document}